\documentstyle[11pt]{article}
\begin{document}
\title{ \bf Explanation of Quantum Mechanics}
\author{Gang Zhao   \\
Department of Physics,Nankai University,Tianjing 300071,P.R.China  \\
E-mail:zhaog@nankai.edu.cn }
\maketitle
\begin{abstract}
 By assuming
 that the kinetic energy,potential energy,momentum,and some other
 physical quantities of a particle exist in the field out of the particle,the Schr$\ddot{o}$dinger
 equation is an equation describing field of a particle,but not the particle itself. \\
 \end{abstract}
 That there are energy and momentum in electromagnetic field is an indisputable fact.
 According to Maxwell equation,density of electromagnetic energy and momentum in vacuum respectively is
 \begin{equation}
 W  =  \frac{1}{2}(\varepsilon_0 E^2 + \frac{1}{\mu_0} B^2)
 \end{equation}
 \begin{equation}
 \bf {p}  =  \varepsilon_0 \bf {E} \times \bf {B}
 \end{equation}
 This means the energy and momentum of electromagnetic field exist in whole space. \\
 The Coulomb interaction of charges can be described in terms of field,too.
 Assume the
 distance of two charges $Q_1,Q_2$ is \bf {\overrightarrow{r}},hence
 \begin{equation}
 \bf {F} = -\nabla U = - \nabla \int \frac{1}{2} \varepsilon_0 E^2 d \tau \nonumber
 \nonumber  \\
 =- \nabla \int \frac{1}{2} \varepsilon_0 E_1\cdot E_2 d \tau =\frac{1}{4 \pi \varepsilon_0}
 Q_1Q_2\frac{\overrightarrow{r}}{r^3}
 \end{equation}

    Note that the potential energy U is an spatial integration,the energy distributes in whole
    space.The energy in any volume element $d\tau$   that contributes to potential energy is
    $ \frac{1}{2} \varepsilon_0 E_1 \cdot E_2 $ ,
    where $E_1,E_2$ are electric amplitudes of $Q_1,Q_2$,respectively.   \\
    Since the electromagnetic energy ,momentum ,potential energy of a particle exist in filed,
    we make a crazy ,but reasonable assumption:the kinetic energy,momentum,potential energy,and
    some other physical quantities of a particle are taken by the field(real field)
     out of the particle(energy distributes in space),but not be taken by the particle itself.
     Hence,if we know the movement equation of field(real field) out of a particle,
     we can calculate the kinetic energy,momentum,potential energy,and some other physical
     quantities of the particle.So we should find the equation of field .The easiest and most
      natural
     way is to assume the Schr$\ddot{o}$dinger equation $i\hbar \frac{\partial}{\partial t}\psi =H\psi$
     is the equation describing the movement of field.    \\
     In classical physics,the equation $\frac{d}{\bf {d} t }\psi =\alpha \nabla^2 \psi$
     can be used as heat field equation[1],that we assume  Schr$\ddot{o}$dinger equation is the
     equation of field is reasonable.\\
     As to the movement of a particle,we can accept L.De Broglie's idea[2] that particle moves
     led by wave,since we admit that particle and field are parts of a body.    \\
     \\
     \bf {References}  \\
     \\
     $1.$ Zhen-Qiu Lu,Equations of Classical  and Modern Mathematical Physics,\\
     Shenhai,1991. \\
     $2.$ L.De.Broglie,La physique quantique restera-telle indeterministe?Gauth
     \\
     ier-Villars,Paric,1953.
     \end{document}